\pdfoutput=1
\documentclass[useAMS,usenatbib,fleqn]{mn2e}
\usepackage{times}
\usepackage{amsmath}
\usepackage{mathrsfs}
\usepackage{amssymb}
\usepackage{amsbsy}
\usepackage{graphicx}
\usepackage{subfigure}
\usepackage{paralist}
\usepackage{color}
\def\df{{\sc df}}
\def\fracj#1#2{{\textstyle\frac#1#2}}
\def\i{{\rm i}}

\def\kpc{\,{\rm kpc}}

\def\T{{\rm T}}

\def\kms{\,{\rm km\,s^{-1}}}

\def\Gyr{\,{\rm Gyr}}

\def\partial{\upartial}
\def\pi{\upi}

\newcommand{\bs}[1]{\bmath{#1}}
\newcommand{\mat}[1]{\mathbfss{#1}}

\newcommand{\e}{\mathrm{e}}
\title[Actions and angles for integrated orbits]{Actions, angles and frequencies for numerically integrated orbits}
\author[J. L. Sanders \& J. Binney]{Jason L. Sanders\thanks{E-mail: jason.sanders@physics.ox.ac.uk} \& James Binney\\Rudolf Peierls Centre for Theoretical Physics, Keble Road, Oxford, OX1 3NP, UK}

\pagerange{\pageref{firstpage}--\pageref{lastpage}} \pubyear{2014}
\label{firstpage}

\begin{document}
\maketitle
\begin{abstract}
We present a method for extracting actions, angles and frequencies from an
orbit's time series. The method recovers the generating function that maps an
analytic phase-space torus to the torus to which the orbit is confined by
simultaneously solving the constraints provided by each time step. We
test the method by recovering the actions and frequencies of tori in a
triaxial St\"ackel potential, and use it to investigate the structure of orbits
in a triaxial potential that has been fitted to our Galaxy's Sagittarius
stream. The method promises to be useful for
analysing $N$-body simulations. It also takes  a step
towards constructing distribution functions  for the triaxial components of
our Galaxy, such as the bar and dark halo.

\end{abstract}


\section{Introduction} 

Although no galaxy is ever in perfect dynamical equilibrium, equilibrium
dynamical models are central to the interpretation of observations of
both our Galaxy and external galaxies. A major reason for the importance of
equilibrium models is that we can infer a galaxy's gravitational potential,
and thus its dark-matter distribution, only to the extent that the galaxy is in
equilibrium. Moreover, equilibrium models are the simplest models and more
complex configurations, involving spiral structure or an on-going minor
merger for example, are best modelled as perturbations of an equilibrium
model.

Globular clusters are the stellar systems that are most completely
understood, and the theory of these systems illustrates the importance of
equilibrium models: at each instant the cluster is assumed to be in dynamical
equilibrium, so, by Jeans' theorem, its distribution function (\df) is a
function of the relevant isolating integrals, such as stellar energy $E$,
total angular momentum $L$, or angular momentum about a symmetry axis, $L_z$.
Over many dynamical times encounters between stars and stellar evolution cause
the \df\ to change, but in such a way that the \df\ continues to satisfy
Jeans' theorem, so the cluster evolves through a series of dynamical
equilibria.

$N$-body simulations of cosmological clustering likewise yield a picture in
which dark-matter haloes are  far from  dynamical equilibrium only during
short-lived and quite rare major mergers. In general a dark-matter halo can be
well approximated by a dynamical equilibrium that is mildly perturbed by
accretion.

The natural way to model a dynamical equilibrium is via Jeans theorem, which
assures us that the system's \df\ can be assumed to be a non-negative
function of isolating integrals. Since one expects a smooth time-independent
gravitational potential to admit up to three functionally independent
isolating integrals, Jeans theorem states that we should be able to represent an
equilibrium stellar system by the density of stars in a three-dimensional
space of integrals rather than in full six-dimensional phase space. This
reduction in dimensionality makes the system very much easier to comprehend
and model.

Since any function of integrals is itself an integral, infinitely many
different integrals may be used as arguments of the \df. However, the action
integrals $J_i$ stand out as uniquely suited to be used as arguments of the
\df. What makes actions special is that they can be complemented by
canonically conjugate variables, the angles $\theta_i$, to form a complete
set of canonical phase-space coordinates. Obviously the equations of motion
of the actions are trivial: $\dot J_i=0$. More remarkably the equations of
motion of the angles are almost as trivial:
$\dot\theta_i=\Omega_i(\bs{J})=\hbox{constant}$. Thus the angle variables
increase linearly in time and if we use angle-action coordinates, the
unperturbed motion of stars becomes trivial. This fact makes angle-action
coordinates uniquely suited to work involving perturbation theory, and indeed
the angle-action coordinates of the Kepler problem were invented to explore
the role played by planet-planet interactions in the dynamics of the Solar
System. 

\cite{McMillanBinney2008} have shown that angle-action coordinates make it possible
to identify stars near the Sun that have been stripped from an object that
was tidally disrupted gigayears ago, and even to determine the date of the
disruption to good precision. \cite{Sellwood11} and \cite{McMillan12} have
used angle-action coordinates to identify stars near the Sun that are
resonantly trapped by spiral structure.  \cite{SandersBinney2013} have shown
how angle-action coordinates for the stars of a stream enable one to
constrain the gravitational potential in which the stream moves.

Cosmological simulations have shown that triaxial dark matter haloes are to be
expected, at least up to the point at which baryons become gravitationally
dominant \citep{Valluri10}. Moreover, \cite{LawMajewski2010} and
\cite{VeraCiroHelmi2013} present evidence that the tidal tails of the
Sagittarius dwarf galaxy can only be fitted if the Milky Way has a triaxial
dark matter halo. Hence we need to be able to determine angle-action
coordinates for stars in triaxial potentials. In this paper we show how to
evaluate the angles and actions of particles in a given triaxial potential.
If the potential is axisymmetric, the actions can be evaluated using the
algorithm given by \cite{Binney12a}.

In Section~\ref{Sec::formal} we derive the equations that yield values of
angles, frequencies and actions. In Section~\ref{Sec::eg} we test our
solutions of these equations by comparing the resulting angles, frequencies
and actions for two orbits in a St\"ackel potential with  analytic values.
In Section~\ref{Sec::appl} we use the equations to explore a constant-energy
surface of the action space of the triaxial potential for our Galaxy that
\cite{LawMajewski2010} fitted to the tidal stream of the Sagittarius dwarf. In
Section~\ref{Sec::discuss} we relate our work to previous work in the field,
discuss a possible extension, and explore how the method copes with resonant
trapping. Section~\ref{Sec::conclude} sums up and
looks to the future.

\section{Formalism}\label{Sec::formal}

Angles and actions can be assigned to orbits that are ``regular'' or
quasiperiodic because such an orbit is confined to a torus labelled by the actions
\citep{Arnold}. We will work in three dimensions so will have three actions
denoted as $\bs{J} = (J_1,J_2,J_3)$. Each  action quantifies the
magnitude of the oscillation in a suitable coordinate. 

The transformation from ordinary phase-space coordinates
$(\bs{x},\bs{v})$ to angle-action coordinates $(\btheta,\bs{J})$ is
possible analytically in only a few cases. \cite{McGillBinney} used one of
these cases as a starting point for the numerical construction of more
general transformations by ``torus mapping''. The key point about torus
mapping is that it yields orbits with specified actions rather than orbits
with specified initial conditions $(\bs{x},\bs{v})$. When analysing an $N$-body
model, we require actions given an initial condition and not vice versa.  Here
we adapt the approach of \cite{McGillBinney} into a procedure which finds the
actions, angles and frequencies given a series of phase-space coordinates
$(\bs{x}_i,\bs{v}_i)$ sampled along an orbit at times $t_i$, where $0\leq t_i\leq T$. 

With this time series we seek a generating function that will map a ``toy
torus'' of a simple ``toy potential'' into the ``target torus'' to which the
orbit is confined. The toy potential must have analytically tractable angles
and actions and permit orbits that have the correct geometry. 

In the absence of figure rotation, a general triaxial potential admits two
basic classes of non-resonant orbit: loop orbits and box orbits
\citep{Schwarzschild79,deZeeuw1985}. Loop orbits have a definite sense of
rotation either around the long- or short-axis of the potential, whilst a box
orbit has no sense of rotation and can reach down to the centre of the
potential.  Hence the class of an orbit can be determined by inspection of
components of the angular momentum along the orbit: if all components of the
angular momentum change sign, the orbit has no sense of circulation and is a
box orbit; when a component of the angular momentum retains its sign,the
orbit is a loop orbit around the corresponding axis \citep{Carpintero98}.
For each class of orbit we use a toy potential that provides tori with the
same geometrical structure as the tori of the given orbit class. 

For a box orbit the actions $J_1$, $J_2$ and $J_3$ quantify the oscillation in
the $x$, $y$ and $z$ directions, respectively.  For loop orbits, $J_1$ quantifies oscillation in a generalized radial coordinate. For a short-axis loop $J_2$ quantifies the particle's circulation around the short axis, whilst $J_3$ quantifies oscillation parallel to this axis. For a long-axis loop orbit, $J_3$ quantifies circulation around the long axis, whilst $J_2$ quantifies oscillation parallel to this axis. We choose this definition such that our actions match $J_\lambda$, $J_\mu$ and $J_\nu$ for a St\"ackel potential \citep{deZeeuw1985}, and each class of orbit occupies a distinct region of action space (see Section~\ref{Sec::appl}).

\subsection{Toy potentials}\label{Sec::toy}

\subsubsection{Triaxial harmonic oscillator}\label{Sec::toy::harmonic}

For box orbits we use the  potential of the triaxial harmonic oscillator,
\begin{equation}
\Phi_{\rm ho}(\bs{x}) = \fracj{1}{2}\sum_{i=1}^3\omega_i^2 x_i^2,
\end{equation}
 which has three parameters, $\omega_i$. Here we have chosen the principal
axes of the potential to lie along the Cartesian $x,y,z$ directions on the
assumption that the time series has already been rotated into the coordinate
system that is aligned with the principal axes of the true potential.  The
actions and angles in this potential are given by
\begin{equation}
\begin{split}
J_i &= \frac{p_i^2+\omega_i^2x_i^2}{2\omega_i},\\
\theta_i &= \arctan\Big(\frac{p_i}{\omega_i x_i}\Big).
\end{split}
\end{equation}

\subsubsection{Isochrone sphere}\label{Sec::toy::isochrone}
For loop orbits we use the isochrone potential,
\begin{equation}
\Phi_{\rm iso}(\bs{x}) = \frac{-GM}{b+\sqrt{b^2+r^2}},
\end{equation}
 where $r$ is the spherical radius. This potential has two free parameters:
the mass $M$ and the scale radius, $b$. The expressions for the actions and
angles in this potential are more involved than for the harmonic oscillator
so are not repeated here. Readers can consult \cite{BinneyTremaine} for the
appropriate equations. The three actions in the isochrone potential are given by the radial action $J_r$, the $z$-component of the angular momentum $L_z$ and the vertical action $J_z\equiv L-|L_z|$, where $L$ is the total angular momentum. With this choice we must orient our coordinate system, such that the orbit circulates around the $z$-axis, before finding the actions.

\subsubsection{Offsets}\label{Sec::toy::offsets}
One might also include the offset of the centre of the potential from the coordinate centre as a free parameter,
but we shall not do so here, presuming instead that the time samples $x_i$
have already been adjusted to be relative to one's best estimate of the
centre of the true potential.

\subsubsection{Parameter choice}
Once a class of potential has been chosen, we set the parameters of the
potential by minimizing \citep{McGillBinney} 
\begin{equation}
\chi^2 = \sum_i (H_i-\langle H\rangle)^2,
\end{equation}
 where the sum is over the times, $H_i$ is the value of the toy Hamiltonian
at $(\bs{x}_i,\bs{v}_i)$, and $\langle H\rangle$ is the mean of these values. The minimization of $\chi^2$ is done using the
Levenberg--Marquardt algorithm \citep{Pressetal}.

The experiments described below suggest that this
method for selecting the parameters is sub-optimal in that it
leads to a rather centrally concentrated toy potential being selected. This
central concentration then leads to high-order Fourier components being
required in the generating function. However, our attempts to find a better
procedure for selecting the toy potential have not met with success.


\subsection{Generating Function}\label{Sec::generatingfn}

With a toy potential chosen, we construct the generating function to
transform between the angle-actions $(\btheta,\bs{J})$ of the toy
potential, and those $(\btheta',\bs{J}')$ of the target potential. The
generating function for this transformation, $S(\btheta,\bs{J}')$, can be
written
\begin{equation}\label{eq:genGF}
S(\btheta,\bs{J}') = \btheta\cdot\bs{J}'
-\i \sum_{\bs{n}\neq\bs{0}}S_{\bs{n}}(\bs{J}')\e^{\i \bs{n}\cdot\btheta},
\end{equation}
 where the vector $\bs{n}$ has integer components. The first term on the right generates the identity transformation, whilst
the structure of the second part is required by the periodicity of the
angle variables. 

\cite{McGillBinney} 
show that if the Hamiltonian is time-reversible, the reality of the
generating function requires the $S_{\bs{n}}$ to satisfy
\begin{equation}
S_{\bs{n}}=-S_{-\bs{n}}.
\end{equation}
  For this condition to be satisfied there must exist a point on the
toy torus at which $\dot{\bs{J}}=\bs{0}$ -- in Appendix~\ref{Appendix::Symmetries} we
demonstrate that this is true for the toy potentials of the previous
section. With this constraint, the generating function can be written as
\begin{equation}\label{eq:GFn}
S(\bs{J}',\btheta) = \btheta\cdot\bs{J}'
+2\sum_{\bs{n}\in\mat{N}}S_{\bs{n}}(\bs{J}')\sin\bs{n}\cdot\btheta,
\end{equation}
 where the integer vectors $\bs{n}$ are now restricted to just half of a
three-dimensional lattice. We take this half to be the set
$\mat{N}=\{(i,j,k)\}$, where either $(k>0)$, $(k=0, j>0)$ or
$(k=0,j=0,i>0)$. Symmetries of the target potential require some of the $S_{\bs{n}}$ to be zero. This is discussed further in Appendix~\ref{Appendix::Symmetries}.

From the generating function (\ref{eq:GFn}) we find that the toy actions are
\begin{equation}
\bs{J}=\frac{\partial S}{\partial \btheta} = \bs{J}'+2\sum_{\bs{n}\in\mat{N}}\bs{n}S_{\bs{n}}(\bs{J}')\cos\bs{n}\cdot\btheta,
\label{toyact}
\end{equation}
and the target angles are
\begin{equation}
\btheta'=\frac{\partial S}{\partial \bs{J}'} 
= \btheta+2\sum_{\bs{n}\in\mat{N}}\frac{\partial S_{\bs{n}}}{\partial \bs{J}'}(\bs{J}')\sin\bs{n}\cdot\btheta.
\label{targetang}
\end{equation}
 Note that by the choice of our generating function, the target angle
zero-point coincides with the toy-angle zero-point.

Given the choice of a toy Hamiltonian, we may find the toy actions and angles
$(\bs{J}(t_i),\btheta(t_i))$ at each time. Each time
then produces a separate equation~\eqref{toyact} with common unknowns:
the target actions and the Fourier components of the generating function,
$S_{\bs{n}}$.

We cannot solve these equations exactly because we are dealing
with equations in an infinite number of unknowns. Because we
can include only a finite number of terms on the right-hand side of each equation,
the right-hand sides should not agree exactly with the left-hand sides, and the correct
procedure is to minimize the sum of the squares of the residuals of
individual equations. This sum is
\begin{equation}
E = \sum_i \sum_k\Big(J_k(t_i)-J'_k-2\sum_{\bs{n}\in\mat{N}}n_kS_{\bs{n}}(\bs{J}')
\cos\bs{n}\cdot\btheta(t_i)\Big)^2,
\end{equation}
 where the inner sum is over the dimension of the action space and the set
$\mat{N}$ is limited to a finite number of vectors $\bs{n}$.  We take this
set to be the $N$ vectors that satisfy the condition
$|\bs{n}|\leq N_{\rm max}$, where $N_{\rm max}\simeq 6$.

We minimize $E$ by setting to zero its derivatives with respect to the
unknowns:
\begin{equation}
\begin{split}
0&=\frac{\partial E}{\partial J_k'} \\
&= -2\sum_i\Big(J_k(t_i)-J'_k-2\sum_{\bs{n}\in\mat{N}}n_kS_{\bs{n}}(\bs{J}')\cos\bs{n}\cdot\btheta(t_i)\Big)\\
0&=\frac{\partial E}{\partial S_{\bs{m}}}\\
&= -2\sum_i\sum_k 2 m_k \cos\bs{m}\cdot\btheta(t_i)\\
 &\times\Big(J_k(t_i)-J'_k-2\sum_{\bs{n}\in\mat{N}}n_kS_{\bs{n}}(\bs{J}')\cos\bs{n}\cdot\btheta(t_i)\Big).
\end{split}\label{eq:basicJ}
\end{equation}

To solve these equations we define a matrix $\mat{c}_{\bs{n}k}$ that has
as subscripts the vector $\bs{n}$ and the integer $k=1,2,3$ that selects a
particular spatial dimension. This $N$-by-$3$ matrix is
\begin{equation}
\mat{c}_{\bs{n} k}(t_i)\equiv 2n_k
\cos\bigl( \bs{n}\cdot\btheta(t_i)\bigr),
\> \,\textrm{(no sum over } \bs{n}\textrm{)}.
\end{equation}
 We further define two $(3+N)$-vectors
\begin{equation}
\bs{x}_{\bs{J}}\equiv (\bs{J}',S_{\bs{n}}),\quad
\bs{b}_{\bs{J}}\equiv \sum_i(\bs{J}(t_i),\mat{c}_{\bs{n}}(t_i)\cdot\bs{J}(t_i)),
\end{equation}
and the symmetric matrix
\begin{equation}
\mat{A}_{\bs{J}} \equiv \sum_i
\left(
\begin{array}{cc}
  \mat{I}_3 & \mat{c}^\T(t_i) \\
  \mat{c}(t_i) & \mat{c}(t_i)\cdot\mat{c}^\T(t_i)\\
\end{array}
\right).
\end{equation}
 Here $\mat{I}_3$ is the 3-by-3 identity matrix. With these definitions,
 the equations (\ref{eq:basicJ}) to be solved can be written as
\begin{equation}
\mat{A}_{\bs{J}}\cdot\bs{x}_{\bs{J}}=\bs{b}_{\bs{J}}. 
\end{equation}
 We solve these equations for $\bs{x}_{\bs{J}}$ by LU
decomposition \citep{Pressetal}.

A similar procedure yields the target angles from equation~\eqref{targetang}. We note that at time $t_i$ the orbit has $\btheta'(t_i) =
\btheta'(0)+\bs{\Omega}'t_i$ where $\bs{\Omega}'$ is the target
frequency, and $\btheta'(0)$ is the angle corresponding to the initial
point in the orbit integration. The relevant sum of squared residuals is
\begin{equation}
F = \sum_i\sum_k\Big(\theta'_k(0)+\Omega'_kt_i-\theta_k(t_i)-2\sum_{\bs{n}\in\bs{N}}\frac{\partial S_{{\bs{n}}}}{\partial J'_k}(\bs{J}')\sin\bs{n}\cdot\btheta\Big)^2.
\label{eq::angpenalty}
\end{equation}
 The unknowns now are $\btheta'(0)$, $\bs{\Omega}'$ and the set of
$\partial S_{\bs{n}}/\partial \bs{J}'$. The requirement of vanishing partial derivatives of $F$ with respect to the unknowns yields the matrix equation,
\begin{equation}
\mat{A}_{\btheta}\cdot\bs{x}_{\btheta}=\bs{b}_{\btheta}.
\end{equation}
These symbols are defined in Appendix~\ref{Appendix::Angles}. The toy angles will be $2\pi$-periodic, and we require the same for the target angles $\btheta'(0)+\bs{\Omega}'t_i$. However, in order to solve the matrix equation we must first make the $\btheta(t_i)$ from the orbit integration continuously increase, and then we solve for the target angles and take the $2\pi$-modulus.

\subsection{Choice of $N_T$, $N_{\rm max}$ and $T$}

Given the scheme presented above, the only questions that remain are how to
select the orbit integration time $T$, the number of time samples, $N_T$, to
use, and what value to use for $N_{\rm max}$, which determines the number $N$
of Fourier components we solve for. Here we discuss how we can automatically
choose these parameters such that we have good recovery of the unknowns.

A necessary condition is that the number of unknowns must be less than the
number of time samples, $N_T$. For the action calculation the number of
unknowns is approximately $N_{\rm max}^3/2$, whilst for the angle-frequency
calculation we have $\sim 3 N_{\rm max}^3/2$ unknowns. We also expect our
ability to recover the unknowns to depend upon the sampling of the toy angle
space.

Let us first consider an idealised 1D case. If we were able to sample
uniformly in the toy angle of a 1D system, we would select $N_T$ points in a
single period separated in toy angle by $\Delta = 2\pi/N_T$. With this
sampling rate we would be able to constrain all modes ${\rm e}^{\i n\theta}$
with $n\Delta\leq\pi$. We can choose to constrain only the $N_{\rm max}$
modes with $n<\pi/\Delta$ as then we would be super-sampling the highest
considered modes. Here we are using a time series that is a product of an
orbit integration so is not uniformly spaced in toy angles -- the toy-angle
distribution depends on the target Hamiltonian, the toy potential and the
distribution of sampling times. The recovery of Fourier components from
non-uniform samples is discussed in \cite{Marvasti2002}. To constrain modes
from a 1D non-uniform sampling we must sample on average at or above the
Nyquist frequency. If we have toy-angle samples $\theta_i$ we require
 \begin{equation}
\frac{n}{N_T-1} \sum_{i=1}^{i=N_T-1} (\theta_{i+1}-\theta_i) \leq \pi,
\end{equation}
to constrain mode $n$.

Here we are attempting to recover components from samples, $\btheta_i$, in 3D
toy-angle space. As we are restricted to using samples generated from an
orbit integration, our sampling is limited to some sub-space of the full 3D
toy-angle space. The 3D sampling can be considered as a series of 1D samples
in $\bs{n}\cdot\btheta_i$ (we first unroll the angles such that they increase
continuously). In order to recover the $S_{\bs{n}}$ from this toy-angle
sampling we need to satisfy two conditions:
\begin{enumerate}
\item As in the 1D case we need to sample on average at or above the Nyquist frequency such that
\begin{equation}
\frac{1}{N_T-1} \sum_{i=1}^{i=N_T-1} \bs{n}\cdot(\btheta_{i+1}-\btheta_i) \leq \pi.
\label{Nyquist3D}
\end{equation}
\item For every included mode, $\bs{n}$, we would also like a good total coverage in $\bs{n}\cdot\btheta$. We choose to require that the $\bs{n}\cdot\btheta$ samples cover the full range from $0$ to $2\pi$:
\begin{equation}
{\rm max}(\bs{n}\cdot\btheta)-{\rm min}(\bs{n}\cdot\btheta)>2\pi.
\label{MinMaxToyAngle}
\end{equation}
If this condition is not satisfied, we are including a mode which will not be well constrained by the toy-angle sampling i.e. the average of $\cos \bs{n}\cdot\btheta$ will not be near zero. We therefore expect that the corresponding $S_{\bs{n}}$ will not be well recovered from this sampling. It could be that this $S_{\bs{n}}$ is not significant so will not affect the recovered actions and frequencies significantly. However, a conservative approach would ensure that equation~\eqref{MinMaxToyAngle} is satisfied for all included modes.
\end{enumerate}
 The second of these conditions is the stricter.  To ensure that the
toy-angle sampling satisfies equation~\eqref{MinMaxToyAngle} when an orbit is
near-resonant, we require time samples which span a very large number of
periods. This is an inevitable drawback of the approach taken here because we
have very little control over the sampling in the toy angle space. 

Having identified a mode which will not be well constrained, one possibility
is to set $S_{\bs{n}}=0$ for this mode. However, by doing this, we risk
throwing out a mode which is significant, and the recovery of the actions and
frequencies will deteriorate so we opt not to do this.

Another requirement is that the $S_{\bs{n}}$ (and $\partial
S_{\bs{n}}/\partial \bs{J}'$) decrease as we go to larger $\bs{n}$ such that
the truncation at $N_{\rm max}$ is valid. If the $S_{\bs{n}}$ do not decrease
with $\bs{n}$, this is evidence of aliasing such that these higher $\bs{n}$
modes are not well recovered and we expect the actions, angles and
frequencies will also not be well recovered. 

\subsubsection{Procedure} 
 
We will now summarize the above discussion into a procedure that can be
implemented:
\begin{itemize}
\item We first select a reasonable $N_{\rm max}$, for instance $N_{\rm max}=6$ is used in the later examples.
\item We then integrate for some time $T$ recording at least $N_T=3N+6$ time samples (or $N_T=N+3$ if we only need the actions) such that we have as many equations as unknowns. This is always satisfied if we choose
\begin{equation}
N_T = {\rm max}(200,\frac{9N_{\rm max}^3}{4}).
\label{NTNMaxrelation}
\end{equation}
\item For each time sample we find the toy angles and check that
equations~\eqref{Nyquist3D} and~\eqref{MinMaxToyAngle} are satisfied for each
mode. If equation~\eqref{Nyquist3D} is not satisfied, $T$ is much longer
than the fundamental orbital periods and so we require a finer time sampling
from the orbit integration. If equation~\eqref{MinMaxToyAngle} is not
satisfied then we continue integrating the orbit until this equation {\em is}
satisfied for all the modes. 

\item We then perform the procedure outlined in
Section~\ref{Sec::generatingfn} to find the $S_{\bs n}$. We require the
$S_{\bs n}$ to be decreasing with $\bs{n}$ such that on the boundaries the
values of the $S_{\bs{n}}$ are small. If we find that the boundary values of
$S_{\bs{n}}$ are large, we have not included a sufficient number of modes in
the generating function so we must increase $N_{\rm max}$ and repeat the
above procedure until we are satisfied that all dominant modes are included.

\end{itemize}
As we will see below this procedure is very conservative but should ensure that the recovery of the actions, angles, frequencies and components of the generating function are accurate.

\section{Example}\label{Sec::eg}

As a test of the above, let us look at an example. The most general separable
triaxial potential is the triaxial St\"ackel potential \citep{deZeeuw1985}. We
choose to work with the perfect ellipsoid, which has density profile
\begin{equation}
\rho(x,y,z) = \frac{\rho_0}{(1+m^2)^2},
\end{equation}
where
\begin{equation}
m^2\equiv\frac{x^2}{a^2}+\frac{y^2}{b^2}+\frac{z^2}{c^2},\>a\geq b\geq c\geq 0.
\end{equation}
The associated coordinates are confocal ellipsoidal coordinates in which
the actions can be expressed as one-dimensional integrals. These may be
calculated numerically using Gauss-Legendre quadrature. Similarly the
frequencies can also be determined from  one-dimensional integrals. Equations for
these quantities are given in \cite{deZeeuw1985}. Here we work with the
potential with parameters $\rho_0 = 7.2\times10^8 M_\odot{\kpc}^{-3},$ $ a =
5.5\kpc,$ $ b = 4.5\kpc$ and $c = 1\kpc$.

\begin{figure}
$$\includegraphics[bb=6 7 240 352]{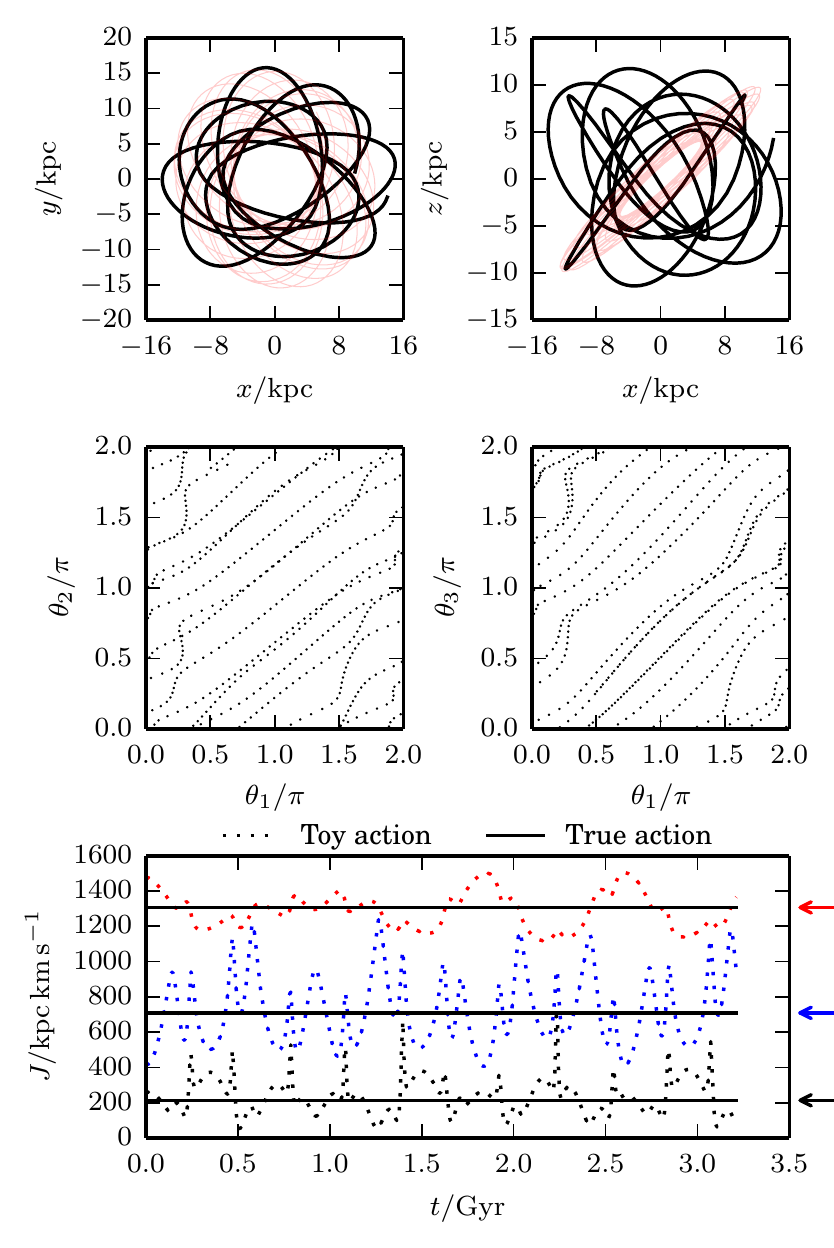}$$
 \caption{Example loop orbit in the triaxial St\"ackel potential -- the top
panels shows in black the orbit integrated in the test potential. This is a
short-axis loop orbit so circulates about the axis $z=0$. In faint red we
show the initial point integrated in the best-fitting isochrone potential. In
the middle panels we show the toy angles calculated at each time sample. In
the bottom panel we show the toy actions at each time-step as a dotted line
(black for $J_1$, blue for $J_2$ and red for $J_3$). The solid lines show the true actions and the arrows mark the estimated actions.}  \label{Loop}
\end{figure}

\begin{figure}
$$\includegraphics[bb=6 7 236 352]{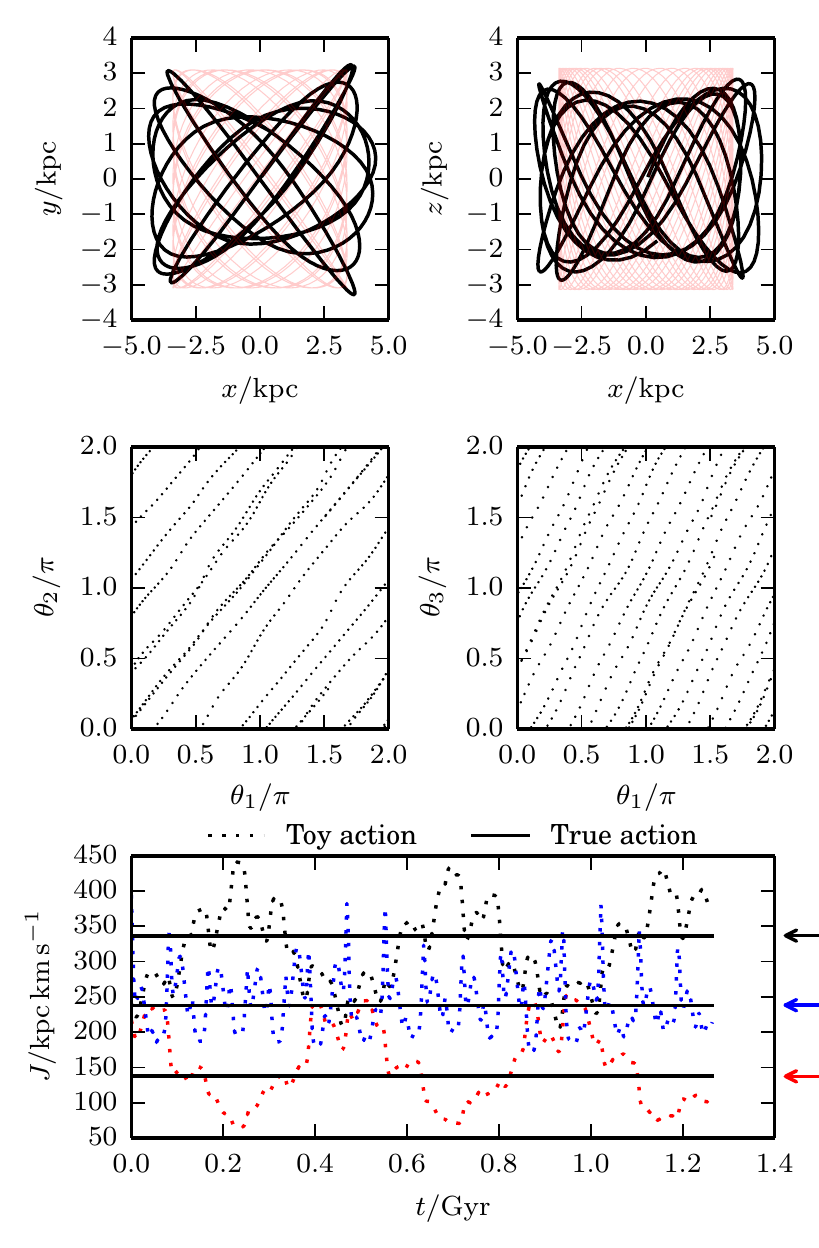}$$
 \caption{As Fig.~\ref{Loop} but for a  box orbit.} \label{Box}
\end{figure}

In this potential we examine two orbits -- a short-axis loop orbit with
initial condition $(x,y,z) = (10,1,8)\,\kpc$, $(v_x,v_y,v_z) =
(40,152,63)\,\kms$ and a box orbit with initial condition $(x,y,z) =
(0.1,0.1,0.1)\,\kpc$, $(v_x,v_y,v_z) = (142,140,251)\,\kms$. Each orbit was
integrated for eight times the longest period, $T_F$. We set $N_{\rm
max}=6$ and calculated the corresponding number of uniformly-spaced time
samples required from equation~\eqref{NTNMaxrelation}. We ensured that
equations~\eqref{Nyquist3D} and~\eqref{MinMaxToyAngle} were satisfied for all
the included modes. In Figs~\ref{Loop} and~\ref{Box} we show the orbits in
the $(x,y)$ and $(x,z)$ planes, the sampling of the toy-angle space and the
resultant actions.  We also show, in faint red, the result of integrating in
the best-fitting toy potential. This gives us an idea of the work that the
generating function has to do to deform the toy torus into the target torus. 

For the loop orbit the true and recovered actions are
\begin{equation*}
\begin{split}
\bs{J}_{\rm true}&=(212.09,1307.54,708.15)\kpc\kms\nonumber\\
\bs{J}_{\rm recov}&=(213.33,1307.29,709.16)\kpc\kms,\nonumber
\end{split}
\end{equation*}
and the true and recovered frequencies are
\begin{equation*}
\begin{split}
\bs{\Omega}_{\rm true}&=(21.76474,15.65172,19.33786)\Gyr^{-1}\nonumber\\
\bs{\Omega}_{\rm recov}&=(21.76508,15.65185,19.33780)\Gyr^{-1}.\nonumber
\end{split}
\end{equation*}

\begin{figure}
$$\includegraphics[bb=33 3 224 245]{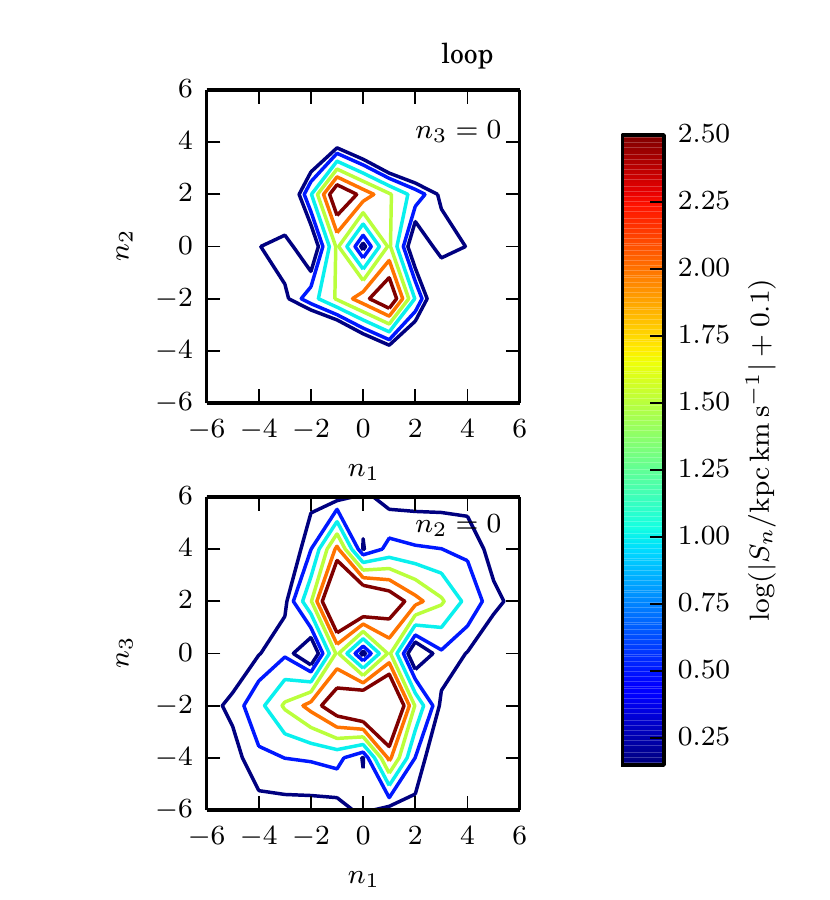}$$
 \caption{Cross-sections of the $S_{\bs{n}}$ as a function of $\bs{n}$ for
the loop orbit. In the top panel we show the cross-section $n_3=0$. The most
significant mode in this plane is $(-1,2,0)$, which causes a mixing between
the radial motion and azimuthal motion. In the lower plane we show the
cross-section $n_2=0$, in which the mode $(0,0,2)$ is the most significant.}
\label{loop_planes}
\end{figure}

\begin{figure}
$$\includegraphics[bb=33 3 224 247]{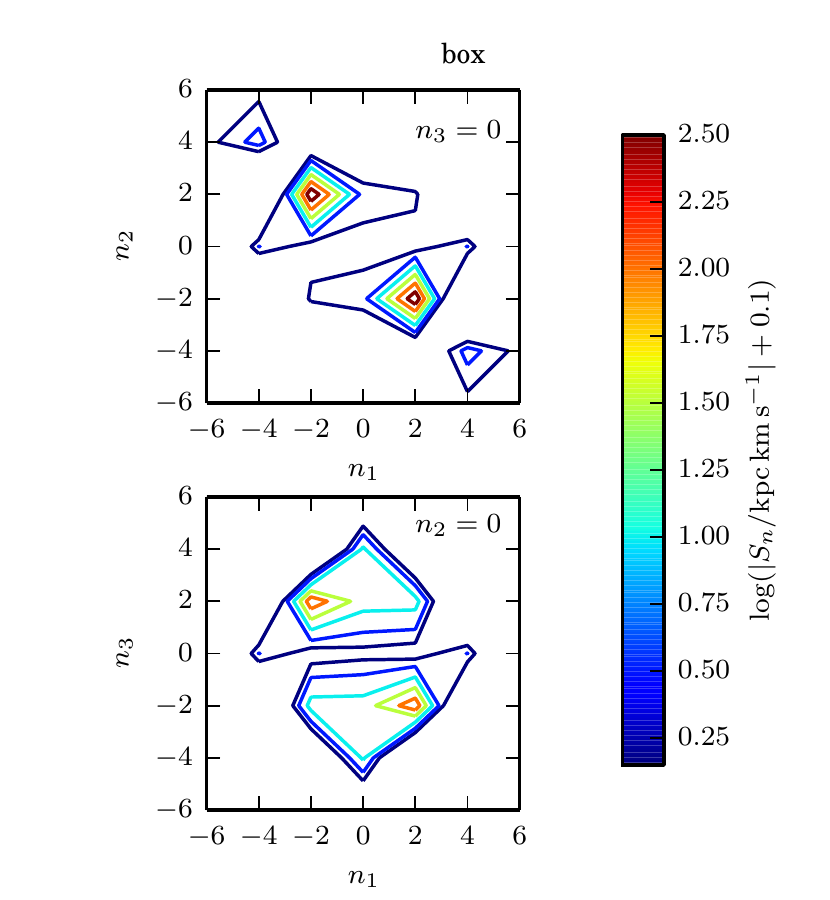}$$
 \caption{As Fig.~\ref{loop_planes} for the box orbit.
The most significant mode in the  plane $n_3=0$ (top) is
$(2,-2,0)$, which causes a mixing between the $x$ motion and the $y$ motion.
In the plane $n_2=0$ the most significant mode is
$(2,0,-2)$.} \label{box_planes}
\end{figure}

In Fig.~\ref{loop_planes} we show two cross-sections of $\bs{n}$-space showing the absolute value of the components of the generating function. (For the isochrone potential we use the convention that subscript 1 refers to $J_r$, subscript 2 refers to $L_z$ and subscript 3 refers to $J_z\equiv L-|L_z|$.) We see that the two most significant modes are $\bs{n} = (-1,2,0)$, which causes a mixing between the radial motion and azimuthal motion, and $\bs{n} = (0,0,2)$. Note that the $S_{\bs{n}}$ decrease towards the boundary so we are content that we have included the relevant modes.

For the box orbit the true and recovered actions are
\begin{equation*}
\begin{split}
\bs{J}_{\rm true}&=(336.39,137.78,237.96)\kpc\kms\nonumber\\
\bs{J}_{\rm recov}&=(336.85,137.26,238.17)\kpc\kms,\nonumber
\end{split}
\end{equation*}
and the true and recovered frequencies are
\begin{equation*}
\begin{split}
\bs{\Omega}_{\rm true}&=(39.752,46.409,73.814)\Gyr^{-1}\nonumber\\
\bs{\Omega}_{\rm recov}&=(39.750,46.406,73.811)\Gyr^{-1}.\nonumber
\end{split}
\end{equation*}

In Fig.~\ref{box_planes} we show two cross-sections of $\bs{n}$-space
showing the absolute value of the components of the generating function.  The
two most significant modes are $\bs{n}=(2,-2,0)$, which causes a mixing
between the $x$ motion and the $y$ motion, and $\bs{n} = (2,0,-2)$, which
mixes the $x$ and $z$ motions. These modes are required to distort the
rectangular orbits of the triaxial harmonic oscillator into those bounded by
surfaces of constant confocal ellipsoidal coordinate. Note that the
$S_{\bs{n}}$ decrease towards the boundaries as required. Also the structure
of Fig.~\ref{loop_planes} is much richer than that of Fig.~\ref{box_planes},
signalling that the
generating function has many more significant terms. 

\subsection{Accuracy of the method}
\begin{figure}
$$\includegraphics[bb=5 8 222 243]{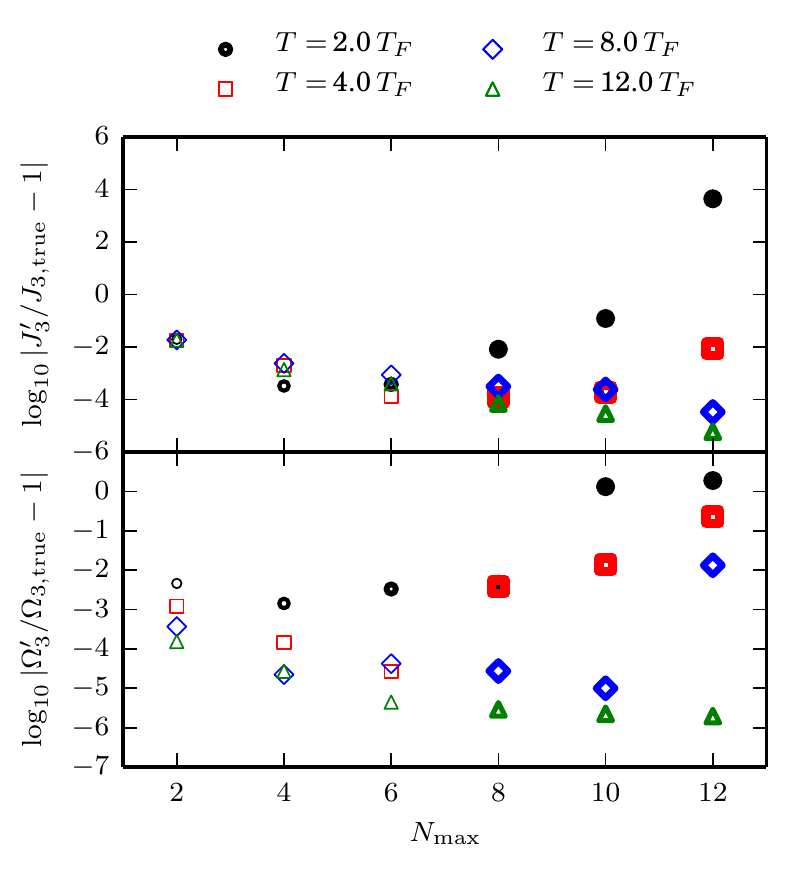}$$
 \caption{Error in the recovered values of  $J_3'$ and $\Omega_3'$ for the box orbit as a function of
$N_{\rm max}$ and the total integration time. We work with even multiples of the period, $T_F$, corresponding to the lowest frequency. The size of the points is proportional to the largest gap in coverage for the $N$ modes. In general a longer integration time provides more accurate actions and frequencies. When attempting to constrain higher modes it is necessary to integrate the orbit for a longer period to ensure that the sampling in toy angle is sufficient.} \label{accuracy}
\end{figure}

Fig.~\ref{accuracy} shows errors in $J_3'$ and $\Omega_3'$ for the box orbit as
a function of $N_{\rm max}$ for various choices of the total integration time $T$. We have linked $N_{\rm max}$ to $N_T$ via equation~\eqref{NTNMaxrelation}. However we have not ensured that equations~\eqref{Nyquist3D} and~\eqref{MinMaxToyAngle} are satisfied for each case. The weight of the points is proportional to the largest gap in coverage for the $N$ modes. We see that in general a longer integration time provides a more accurate estimate of the action and particularly the frequency. We can understand this as a longer line segment provides a better measurement of the gradient for noisy data. From Fig.~\ref{accuracy} we see that when working with high $N_{\rm max}$ it is not sufficient to satisfy equation~\eqref{NTNMaxrelation}. We must also satisfy equation~\eqref{MinMaxToyAngle} such that we have a sufficient sampling in toy-angle space to constrain these higher modes. 

For $T=2T_F$ equation~\eqref{MinMaxToyAngle} is not satisfied for $N_{\rm
max}\geq4$. For large $N_{\rm max}$ and $T=2T_F$ many modes have insufficient
coverage and the results are very poor. For the other three integration times
equation~\eqref{MinMaxToyAngle} is not satisfied for $N_{\rm max}\geq8$. For
$T=4T_F$ this results in an immediate deterioration of the frequency recovery
as we have included a mode with ${\rm max}(\bs{n}\cdot\btheta)-{\rm
min}(\bs{n}\cdot\btheta)\approx\pi/2$. For $T=8T_F$ and $T=12T_F$ a lack of
coverage has not affected the results apart from for $T=8T_F$ and $N_{\rm
max}=12$ where the frequency recovery is poorer. The mode which is not well
covered is also not well covered for $N_{\rm max}=8$ but we only see the
effects of this lack of coverage when we try to include more modes. For
$T=12T_F$ both the action and frequency recovery are very good despite
equation~\eqref{MinMaxToyAngle} not being satisfied when $N_{\rm max}\geq 8$.
In particular there is one mode for which ${\rm max}(\bs{n}\cdot\btheta)-{\rm
min}(\bs{n}\cdot\btheta)\approx4.3$. It seems that this coverage is
sufficient to not degrade the results. In conclusion, when
equation~\eqref{MinMaxToyAngle} is satisfied we recover the frequencies and
actions well, whilst when it is not satisfied the recovery deteriorates in
some cases, particularly that of the frequency. 

Finally, we find that when we double the number of time samples used for the
examples shown in Fig.~\ref{accuracy} the results change significantly
only when equation~\eqref{MinMaxToyAngle} is not
satisfied. Therefore, we conclude that provided we have more equations than
unknowns and have satisfied equations~\eqref{Nyquist3D}
and~\eqref{MinMaxToyAngle} the actions and frequency recovery will be
satisfactory.

\subsection{Near-resonant orbit}
\begin{figure}
$$\includegraphics[bb=6 7 240 352]{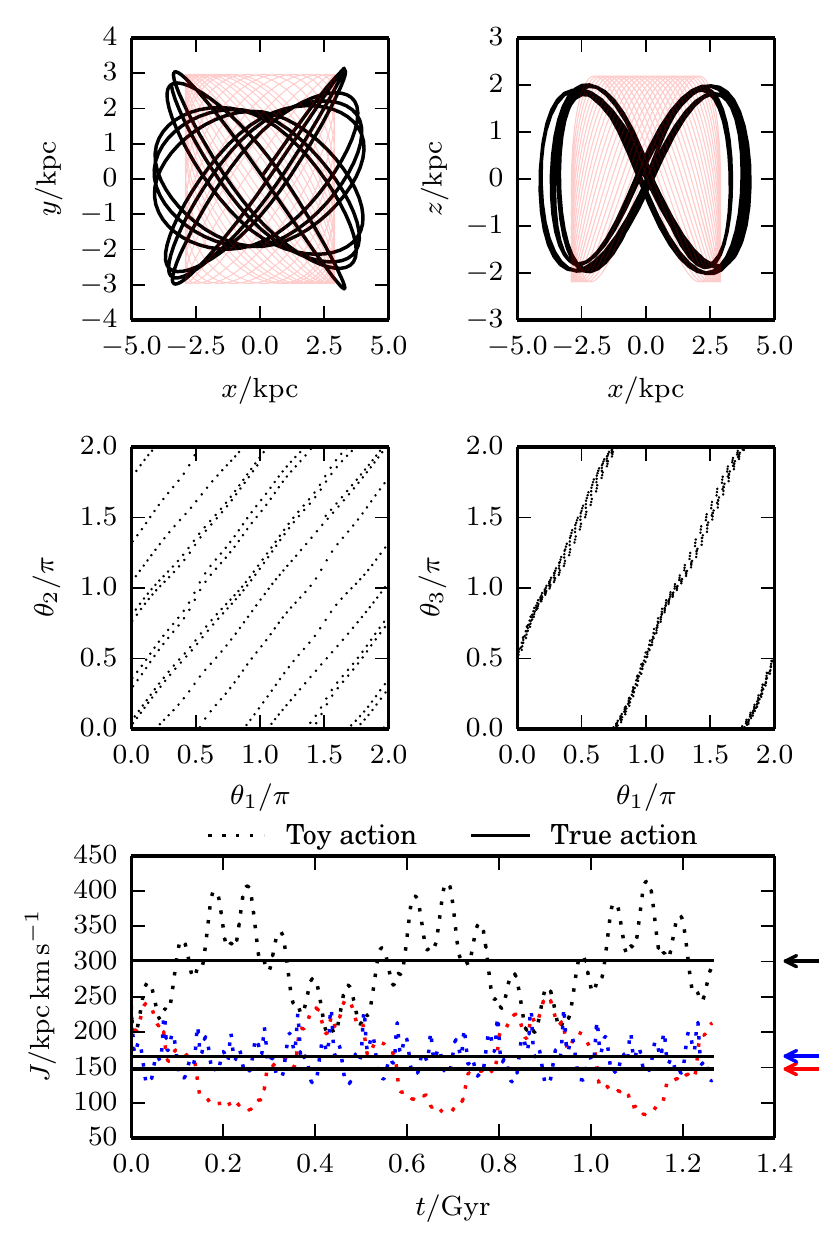}$$
 \caption{Example near-resonant orbit in the triaxial St\"ackel potential -- the top
panels shows in black the orbit integrated in the test potential. This is a
short-axis loop orbit so circulates about the axis $z=0$. In faint red we
show the initial point integrated in the best-fitting isochrone potential. In
the middle panels we show the toy angles calculated at each time sample. In
the bottom panel we show the toy actions at each time-step as a dotted line
(black for $J_1$, blue for $J_2$ and red for $J_3$). The solid lines show the true actions and the arrows mark the estimated actions.}  \label{Res}
\end{figure}

To illustrate some of the points discussed we show results for a
near-resonant orbit. This orbit is a box orbit with the initial conditions
$(x,y,z) = (0.1,0.1,0.1)\,\kpc$, $(v_x,v_y,v_z) = (142,150,216.5)\,\kms$.
Again we integrate for time $T=8T_F$ and set $N_{\rm max}=6$. The results are
shown in Fig.~\ref{Res}. The frequency vector of this orbit is nearly
parallel to $\bs{n} = (-4,0,2)$ so the coverage of this mode is very poor and
${\rm max}(\bs{n}\cdot\btheta)-{\rm min}(\bs{n}\cdot\btheta)\approx1.11$ for
$\bs{n} = (-4,0,2)$. However, the true and recovered actions are
\begin{equation*}
\begin{split}
\bs{J}_{\rm true}&=(301.74,147.63,165.36)\kpc\kms\nonumber\\
\bs{J}_{\rm recov}&=(300.69,147.66,165.89)\kpc\kms,\nonumber
\end{split}
\end{equation*}
and the true and recovered frequencies are
\begin{equation*}
\begin{split}
\bs{\Omega}_{\rm true}&=(43.318,50.369,86.724)\Gyr^{-1}\nonumber\\
\bs{\Omega}_{\rm recov}&=(43.386,50.371,86.777)\Gyr^{-1}.\nonumber
\end{split}
\end{equation*}
As seen before, poor coverage in one of the modes is not detrimental to the action and frequency recovery.

\section{Application}\label{Sec::appl}

As a brief application of the method outlined in this paper we will inspect
the action diagram for a realistic triaxial Galactic potential. We take the
potential from \cite{LawMajewski2010}. This potential was found to produce
the best fit to the Sagittarius stream data. This potential has three
components: a disc defined  by the Miyamoto-Nagai potential
\begin{equation}
\Phi_{\rm disc}(x,y,z) = \frac{-GM_{\rm
disc}}{\sqrt{x^2+y^2+(a+\sqrt{z^2+b^2})^2}},
\end{equation}
 with $M_{\rm disc} = 10^{11}\, M_{\odot}$, $a=6.5\kpc$ and $b=0.26\kpc$; a
spherical bulge described by the Hernquist profile
\begin{equation}
\Phi_{\rm bulge}(r) = \frac{-GM_{\rm bulge}}{r+c},
\end{equation}
 with $M_{\rm bulge} = 3.4\times 10^{10}\, M_{\odot}$ and $c=0.7\kpc$; and the
triaxial logarithmic halo
\begin{equation}
\Phi_{\rm halo}(x,y,z) = v_{\rm halo}^2\log\Big(C_1x^2+C_2y^2+C_3xy
+\frac{z^2}{q_z^2}+r_{\rm halo}^2\Big)
\end{equation}
 with $v_{\rm halo} = 121.7\kms$, $C_1 = 0.99\kpc^{-2}$, $C_2=0.53\kpc^{-2}$,
$C_3 = 0.11\kpc^{-2}$, $q_z=1.36$ and $r_{\rm halo} = 12\kpc$.

\subsection{An example orbit} 

\begin{figure}
$$\includegraphics[bb=6 7 231 352]{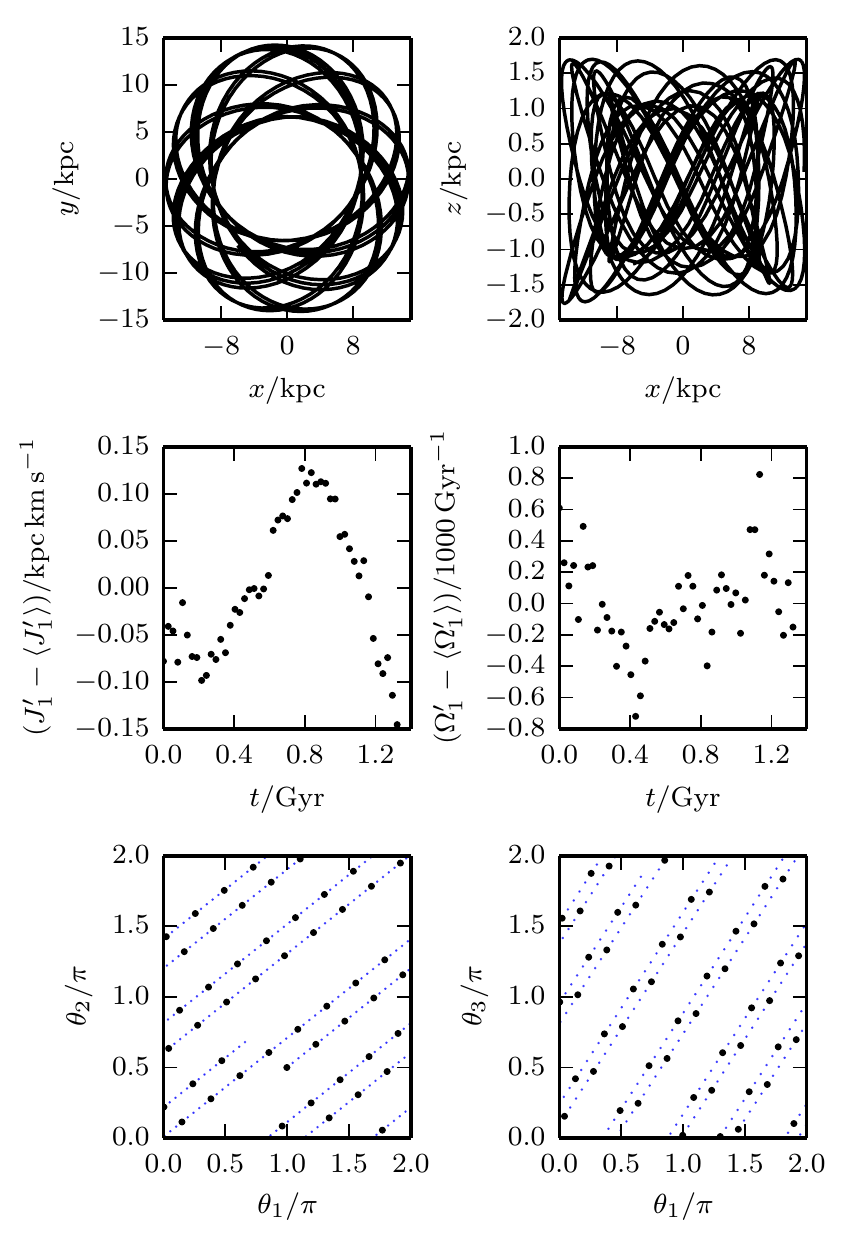}$$
 \caption{An example orbit in the \protect\cite{LawMajewski2010} potential.
It is a short-axis loop orbit with actions $\bs{J}' \approx (160, 2186,
36)\kpc\,\kms$. In the top panel we show a $16T_F$ long orbit segment in
the $(x,y)$ and $(x,z)$ planes. In the central two panels we show the spread in $J_1'$ and $\Omega_1'$ calculated using $500$ time-samples from an $8T_F$ orbit segment labelled by its initial time sample. In the bottom panel we show the calculated angles at these times with black dots. We also show the angles found using $\btheta'(0)+\bs{\Omega}'t_i$ with one of the calculated frequencies and initial angles in smaller blue dots.} \label{LMpot_test}
\end{figure}

We inspect a single orbit in this potential in Fig.~\ref{LMpot_test}. The
chosen orbit is a short-axis loop orbit with initial condition $(x,y,z) =
(14.69,1.80,0.12)\,\kpc$, $(v_x,v_y,v_z) = (15.97,-128.90,44.68)\,\kms$.  We
use different, but overlapping, $8T_F$ long segments of the orbit with
$N_T=500$ to calculate the actions, angles and frequencies using $N_{\rm
max}=6$. We ensure equation~\eqref{MinMaxToyAngle} is satisfied for these time
samplings. This orbit lies in the surface of constant energy explored in the
next section. We find that the action and frequency are
\begin{eqnarray}
\bs{J}' &\approx& (160.18,2186.16,36.09)\kpc\,\kms\nonumber\\
\bs{\Omega}'&\approx&(27.26,19.12,37.01)\Gyr^{-1}.\nonumber
\end{eqnarray}
The error in the actions and frequencies can be estimated by the spread of
the estimates from each segment. We find
\begin{eqnarray}
\Delta\bs{J}'&\approx&(0.07,0.08,0.03)\kpc\kms,\nonumber\\
\Delta\bs{\Omega}'&\approx&(3\times10^{-4}, 6\times 10^{-5},2\times 10^{-3})\Gyr^{-1}.\nonumber
\end{eqnarray}
For each orbit segment we find $\btheta'(0)$ and these different values
should  all lie along straight lines with gradients given by the derived frequencies. In Fig.~\ref{LMpot_test} we show that the condition is well satisfied.

Using different orbit segments is perhaps the only way to estimate the error in an action or frequency found using the present method. It is simplest to use consecutive orbit segments as we have here. However, a better method is to use orbit segments separated by a large time interval. This can be achieved most effectively by utilizing the estimated generating function to find an initial condition for a second orbit integration. A simple choice is to increase one of the derived angle coordinates by $\pi/2$. 

\subsection{A typical constant energy surface} 

\begin{figure}
$$\includegraphics[bb=15 12 247 366]{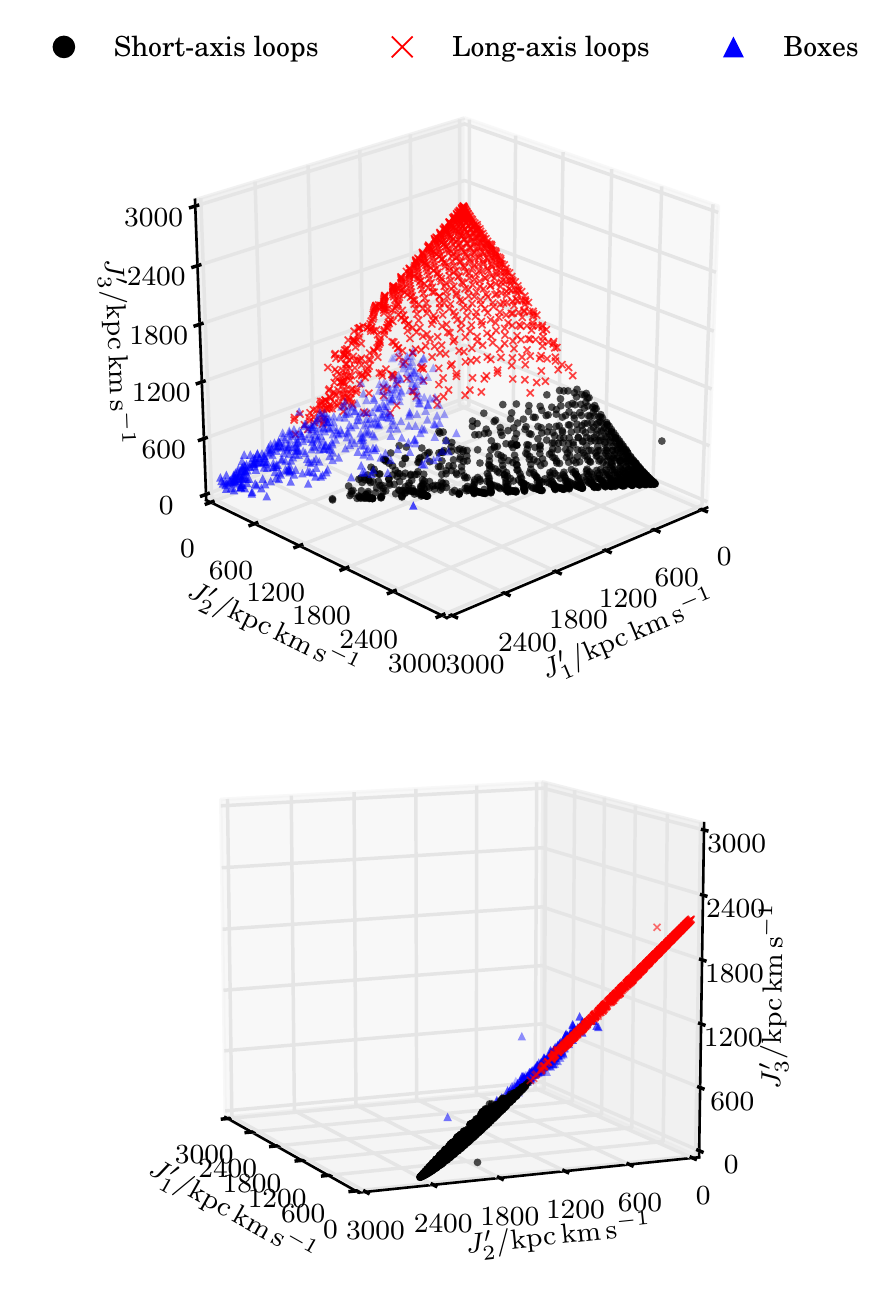}$$
\caption{Two projections of a surface of constant energy in the 3D action space of the potential proposed by \protect\cite{LawMajewski2010}. Black circles show short-axis loop orbits, red crosses show long-axis loop orbits and blue triangles show box orbits.}
\label{LMpot}
\end{figure}

Now we turn to constructing the action diagram for the chosen potential. For
a given energy (that of a particle dropped at $18\kpc$ on the intermediate
axis) we launched particles at a series of points linearly spaced between
$0.2$ and $18\kpc$ along the potential's intermediate axis with the
velocity vector perpendicular to the axis and inclined at linearly spaced
angles to the $z$-axis between $0$ and $\pi/2$\footnote{Note that the
intermediate axis of the halo model proposed by \citeauthor{LawMajewski2010}
is actually the $z$-axis. However, at small radii ($\lesssim 18\kpc$) the
intermediate axis of the full potential is in the $(x,y)$ plane due to the
disc contribution, and the $z$-axis is the short axis.}. We integrated each
initial condition for $\sim10\Gyr$ saving $N_T=1000$ samples. For all orbits
the energy was conserved to one part in $10^6$. We set $N_{\rm max}=6$ and
ensured that equations~\eqref{Nyquist3D} and~\eqref{MinMaxToyAngle} were
satisfied. If equation~\eqref{Nyquist3D} was not satisfied, we had
undersampled the orbit, so we took a finer sampling. If
equation~\eqref{MinMaxToyAngle} was not satisfied we did not have sufficient
coverage, so we continued integrating for another $10\Gyr$, taking another
$1000$ samples. We then calculated the actions from the time series.
Fig.~\ref{LMpot} shows each orbit as a point in 3D action-space\footnote{To
produce a continuous plane in action-space we must scale the `radial' actions
of the loop orbits, $J_1$, by a factor of $2$. $J_1$ for a loop orbit
corresponds to a single oscillation from minimum to maximum coordinate and
back, whilst for a box orbit a single oscillation covers the interval $0$ to
maximum coordinate four times.}. We see that the surface of constant energy
is a triangle-shaped plane in action-space. The points are coloured based on
their orbit classification. An equivalent figure for a St\"ackel potential
can be found in \cite{deZeeuw1985}.

In a triaxial potential, the loop orbits can be divided into two classes: the
short-axis loops that loop around the short axis (in our case the $z$-axis)
and the long-axis loops that loop around the long axis (the $x$-axis). Along
with the box orbits these three classes of orbit occupy distinct regions on
the action-space plane of constant energy. At each corner of the plane only
one action is non-zero and the corresponding orbit is the parent orbit of
each of the three classes: the $J_2=0,J_3=0$ orbit is a radial orbit along
the long axis, the $J_1=0,J_3=0$ orbit is a closed orbit in the $(x,y)$ plane
and the $J_1=0,J_2=0$ orbit is a closed orbit in the $(y,z)$ plane. We note
that near the interface between the different orbit classes some regions of
the plane are depleted of points (our choice of initial sampling causes an
increased density of points near the edges of the plane). Also there is some
overlap between the different orbit classes in the action space. These
features are due to the presence of resonant islands with surrounding chaotic
orbits at the interface of the regular orbit regions (see
Section~\ref{Sec::chaos}). For orbits near the box/loop boundary it can take
many orbital periods to correctly identify the orbit class
\citep{Carpintero98}, and some may be misclassified.

\section{Discussion}\label{Sec::discuss}

\subsection{Relation to previous work}

The problem addressed here goes back to \cite{BinneySI,BinneyS}, who Fourier
transformed the time series $x(t_i)$ of individual coordinates and assigned
to each line in the resulting spectrum appropriate integers $n_j$ so that
$\omega t$ could be identified with $\sum_j n_j\Omega_jt$. Once this
identification had been successfully accomplished, $\Omega_jt$ could be
replaced with $\theta_j$ to yield the orbit's angle representation. This
approach is inferior to that introduced here in several respects: (i) Whereas
the generating function is a scalar, a star's location is described by a
vector, so it is wasteful to construct the angle representations of all three
coordinates rather than the angle representation of the generating function:
\cite{BinneyS} failed to take advantage of the strong restrictions on tori
that arise from angle-action coordinates being canonical. (ii) It is not
straightforward to measure correctly the complex amplitudes $A$ from the
discrete Fourier transform of a time series such as $x(t_i)$ because the
required amplitude will in general not lie at one of the discrete frequencies
sampled. (iii) When an orbit is near-resonant there is often dangerous
ambiguity in the integers $n_j$ that should be assigned to a particular line.
With the present technique we work from the outset with periodic functions
and their Fourier series so the issue of how frequencies fall on a discrete
grid does not arise. Moreover, the assignment of integers $n_j$ to Fourier
terms is unambiguous.

The method described here has significant overlap with the work of
\cite{Warnock} on the construction of magnetic coordinates and the related
method of \cite{KaasalainenB} for the construction of angle coordinates. In
both these studies angle-action variables were evaluated along numerically
computed orbits. The coordinates evaluated were not those of a toy potential
but of a trial torus that had been previously constructed: \cite{Warnock} was
refining the Fourier coefficients $S_{\bs{n}}$ while \cite{KaasalainenB} were
solving for the $\partial_iS_{\bs{n}}$ given the $S_{\bs{n}}$. In both these
studies, several initial conditions for orbit integration were chosen on each
torus to overcome the problem that with a single short integration a resonant
orbit yields a highly non-uniform distribution of sample points on the torus.
Since we do not have a good representation of the target torus until the
equations have been set up and solved, we cannot take advantage of this
possibility.

\cite{Warnock} solved for the discrete Fourier transforms of the
$\bs{n}S_{\bs{n}}$ rather than for the $S_{\bs{n}}$ because the matrix that then
has to be inverted is nearly diagonal when the toy and target tori are close
to one another and the sample points provide a nearly regular grid in the
space of toy angles. Since our toy and target tori can be quite different,
and it is hard to achieve a uniform sampling of toy-angle space, we have not
used Warnock's technique. 

\subsection{Possibility of using St\"ackel tori}

We have used completely different toy potentials for each class of orbit, and
it is natural to ask whether it would not be advantageous to use always a
St\"ackel potential since such a potential has tori of every type. We have
not pursued this option for two reasons. First, the actions and angles of
St\"ackel potentials require the evaluation of integrals whereas the
potentials we have used yield algebraic expressions for angles and actions.
Secondly, and more fundamentally, when integrating an orbit that lies close to
the box/loop interface, it would be non-trivial to ensure that the toy torus
with the actions of the target orbit had the same geometry as the target
torus. By using potentials that support only one type of torus, we are
assured from the outset that this condition is satisfied. However, this rests on our correct identification of the orbit type from the time series. As we saw with the \citeauthor{LawMajewski2010} potential, in some marginal cases it may take many orbital periods to correctly identify the orbit.

\subsection{Resonances and chaos}\label{Sec::chaos}

We have focused here on orbits that are non-resonant members of the major
orbital families. In real galactic potentials one encounters orbits that are
either resonantly trapped or chaotic \citep[e.g.\ \S3.7][]{BinneyTremaine}.
Chaotic orbits can be thought of as sequences of sections of resonantly
trapped orbits, so these two types of orbit raise similar issues.

In a generic integrable potential, the frequencies $\Omega_i$ depend on the
actions, so on some tori a resonant condition $\bs{n}\cdot\bs{\Omega}=0$ is
satisfied.  Consequently, individual orbits on these resonant tori do not
cover the entire torus since the condition
$\bs{n}\cdot\btheta=\hbox{constant}$ constrains the angle variables. This
lack of coverage makes it impossible to determine some of the Fourier
coefficients $S_{\bs{n}}$. 

When the potential is strictly integrable, orbits on tori that are adjacent
to a resonant torus completely cover their tori although they take a long
time to do so. In a generic potential, however, such orbits move over a
series of tori without covering any of them, as they librate around the
strictly resonant orbit.  Consequently, these orbits have some of the
characteristics of a strictly resonant torus. When the present technique is
used on a resonantly trapped orbit, the generating function will map the toy torus into a close approximation to the strictly resonant torus, so in an
$N$-body model the density of stars on this torus will seem to be larger than
it really is.  Hence with the present technique, resonantly trapped orbits
will give rise to apparent crowding in action space that is analogous to the
signature of resonances when particles are mapped into frequency space by
determining orbital frequencies by Fourier decomposition of coordinates
\citep{Laskar}: when the ratios $\Omega_2/\Omega_1$ and $\Omega_3/\Omega_1$
are used to place orbits in frequency-ratio space, the existence of
resonantly trapped orbits leads to a crowding of points along the straight
lines associated with certain resonance conditions
$\bs{n}\cdot\btheta=\hbox{constant}$ \cite[\S 3.7.3(b)]{BinneyTremaine}.

Chaotic orbits can be considered as moving through a series of quasi-periodic orbits. Therefore the recovered actions and frequencies from our method will be a function of the total integration time. We see that the region of the constant energy surface occupied by the box orbits in Fig.~\ref{LMpot} has considerable crowding and the regular grid of initial conditions is not visible. This is indicative of chaotic orbits which have been allocated very different actions from one initial condition to the next.

\begin{figure}
$$\includegraphics[bb=6 7 231 250]{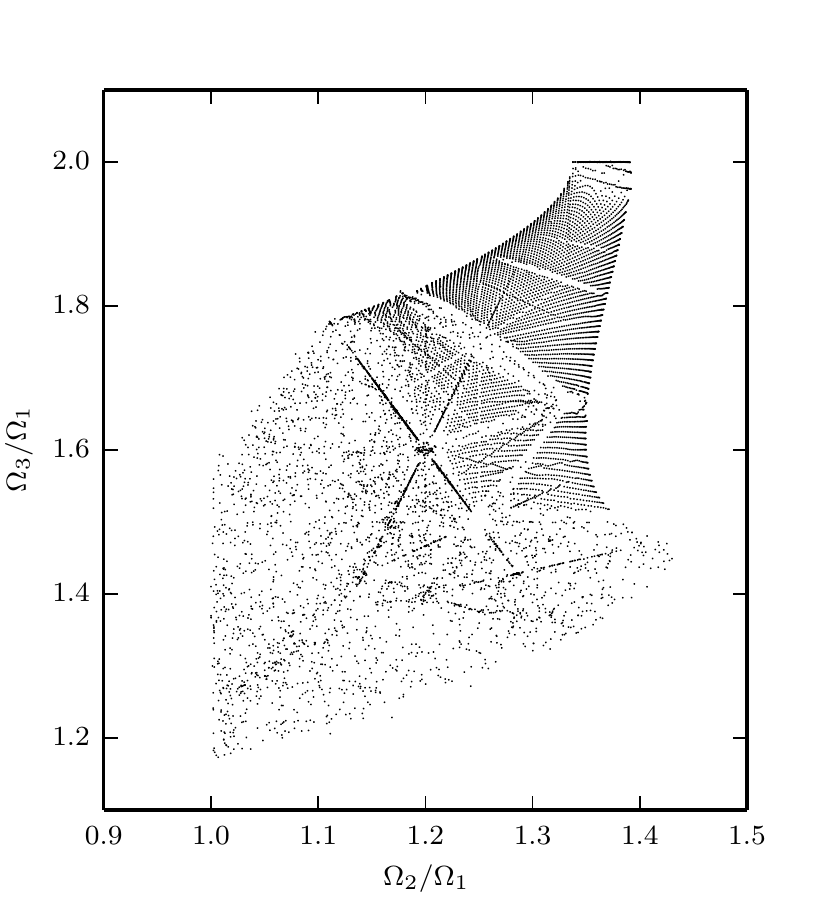}$$
 \caption{Frequency ratios in the triaxial logarithmic potential extracted from orbital time series using the method presented in this paper. Each point corresponds to an initial condition for a particle dropped on the surface $\Phi(\bs{x})=0.5$.} \label{BinneyTremaineF}
\end{figure}

In Fig.~\ref{BinneyTremaineF} we perform the same procedure as outlined in \S 3.7.3(b) of \cite{BinneyTremaine} to inspect the ratio of frequencies plane of a logarithmic potential. We use the potential
\begin{equation}
\Phi(\bs{x}) = \frac{1}{2}\log\Big(x^2+\frac{y^2}{q_y^2}+\frac{z^2}{q_z^2}+r_c^2\Big),
\end{equation}
with $q_y=0.9$, $q_z=0.7$ and $r_c^2=0.1$. We drop a series of test particles
on the surface $\Phi(\bs{x})=0.5$ regularly spaced in the spherical polar
coordinates $\phi$ and $\cos\theta$, and integrate each initial condition for
a time $T=200$ extracting $N_T=2048$ samples. We then use our method to find the
corresponding orbital frequencies and plot their ratio in
Fig.~\ref{BinneyTremaineF}. As noted in \cite{BinneyTremaine} the top-right
corner of this plane shows the regular spacing of the initial conditions
whilst the lower-left corner shows a more irregular distribution with no
evidence of the regular grid of initial conditions used to produce it. Also,
we find that there are overdensities along lines corresponding to resonances.
Our plot is very similar to that shown in \cite{BinneyTremaine}. However, the
structure of the irregular bottom-left region differs. This is to be expected
as it is these orbits which are irregular, and how one assigns regular
properties to them depends on the method employed.

\section{Conclusions}\label{Sec::conclude}

We have presented a method for finding actions, frequencies and angles from
numerically integrated orbits in a general potential\footnote{
We will make the code developed for this paper available at 
{\tt https://github.com/jlsanders/genfunc}.}. The method relies on
estimating the Fourier components of the generating function that maps a toy
torus into the torus on which the computed orbit lies by solving systems of
linear algebraic equations. This method enables one to determine the
angle-action coordinates $(\btheta,\bs{J})$ of a given phase-space point
$(\bs{x},\bs{v})$. It has numerous possible applications in astronomy.

Ours is the first method presented in the literature for finding the actions
in a general triaxial potential. Triaxiality is an essential ingredient of
dark-matter distributions, and a realistic Galactic model which should include
non-axisymmetric features such as the bar, and the potentially triaxial halo.
This method is a necessary first step towards constructing distribution
functions, $f(\bs{J})$, for these more complex Galactic components.

An important application is to the analysis of $N$-body simulations.  A
single $N$-body snapshot consists of 3D positions and velocities for $\sim10^9$
particles. Letting the simulation evolve for a few time steps produces
another snapshot with a completely different set of $10^9$ positions and
velocities. Thus the particles' phase-space coordinates constitute a highly
degenerate and non-compact representation of the simulation.  Effective
analysis of the simulation should start by condensing the coordinates into a
smaller set of numbers. This can be done by replacing the $6N_T$ numbers
$(\bs{x}_i,\bs{v}_i)$ with just three numbers $J_i$ and plotting each
particle as a point in 3D action space. The simulation then becomes a density
of particles in a 3D space. This representation will greatly facilitate the
comparison of different $N$-body models. Also it may prove possible to find
good fits to the star density in terms of analytic functions, as
\cite{Pontzen} have done for numerical dark-matter haloes and appears to be
possible for the Galactic discs \citep{Binney12b,Binney2013}.  We hope to
report on an application of this method to an $N$-body simulation soon.

It should be noted that it is not advisable to take the $N_T$ time samples of
a given orbit directly from the simulation. Rather at some time $t$ the
potential should be computed on a spatial grid \citep[e.g.][]{Magorrian07}, and
the equations of motion in this potential should be integrated for $N_T$
timesteps starting from the phase-space location of each particle at time
$t$. These integrations in a fixed potential lend themselves to massive
parallelization, for example on a Graphical Processor Unit (GPU) so it should
be possible to compute angle-action coordinates for very large numbers of
particles .

Here we discussed time-reversible triaxial potentials. In this case we can
determine a priori the phases of the terms in the generating function.
Rotation of the figure of the potential destroys the time-reversibility of
the Hamiltonian and we lose the ability to set the phases a priori.  In the
worst case, the $S_{\bs{n}}$ in equation (\ref{eq:genGF}) become complex
numbers that are only limited by the condition $S_{-\bs{n}} = S_{\bs{n}}^*$
required to make the generating function real. Extending the current
framework to this case $\sim$ doubles the dimensionality of the matrices we
must solve for given $N_{\rm max}$.

\section*{Acknowledgements} JS acknowledges the support of STFC. JB was
supported by STFC by grants R22138/GA001 and ST/K00106X/1. The research
leading to these results has received funding from the European Research
Council under the European Union's Seventh Framework Programme
(FP7/2007-2013) / ERC grant agreement no.\ 321067.

{\footnotesize{
\bibliographystyle{mn2e}
\bibliography{genfunc_bib}
}}

\appendix
\section{Symmetries}\label{Appendix::Symmetries}

In Section \ref{Sec::generatingfn} we asserted that for a time-reversible Hamiltonian the
Fourier components of the generating function, $S_{\bs{n}}$, are real.
However, it must also be true that there is a point on the target torus where
$\dot{\bs{J}}=\bs{0}$.  \cite{McGillBinney} show that this is true if the toy
potential is an isochrone and the target Hamiltonian is axisymmetric.
Additionally they demonstrated that when the potential is symmetric about the
plane $z=0$, Fourier components of the generating function with odd $n_z$
vanish. Here we repeat these arguments extended to the 3D
triaxial case. 
\subsection{Loop orbits}
Let us first consider the loop orbits. Suppose we have a target
Hamiltonian of the form
\begin{equation}
H(r,\phi,\vartheta) = \fracj{1}{2}p^2_r+\frac{p^2_\phi}{2r^2\sin^2\vartheta}+\frac{p^2_\vartheta}{2r^2}+\Phi(r,\phi,\vartheta),
\end{equation}
where $(r,\phi,\vartheta)$ are standard spherical polar coordinates. The equations of motion for the toy actions are
\begin{equation}
\begin{split}
\dot{J}_i = &-\frac{\partial H}{\partial \theta_i}\\=&\Big(\frac{p^2_\vartheta}{r^3}+\frac{p^2_\phi}{r^3\sin^2\vartheta}-\frac{\partial\Phi}{\partial r}\Big)\frac{\partial r}{\partial \theta_i}+\Big(\frac{p^2_\phi\cos\vartheta}{r^2\sin^3\vartheta}-\frac{\partial\Phi}{\partial \vartheta}\Big)\frac{\partial \vartheta}{\partial \theta_i}\\&-\frac{\partial\Phi}{\partial\phi}\frac{\partial\phi}{\partial\theta_i}-p_r\frac{\partial p_r}{\partial \theta_i}-\frac{p_\vartheta}{r^2}\frac{\partial p_\vartheta}{\partial \theta_i}-\frac{p_\phi}{r^2\sin^2\vartheta}\frac{\partial p_\phi}{\partial \theta_i}.
\end{split}
\end{equation}
 Now let us consider the point $\btheta=(0,0,\pi/2)$: at this point the
particle is at pericentre, at a maximum in its vertical oscillation and at
$\phi=0$. Therefore at this point we have that
\begin{equation}
\frac{\partial r}{\partial \theta_i} = \frac{\partial \vartheta}{\partial \theta_i} = p_r = p_\vartheta = \frac{\partial p_\phi}{\partial \theta_i} = 0,
\end{equation}
so
\begin{equation}
\dot{J}_i = -\frac{\partial\Phi}{\partial\phi}\frac{\partial\phi}{\partial\theta_i}.
\end{equation}
 In a triaxial potential with its axes aligned with the coordinate axes, $x=0$ is a symmetry plane of the potential so
$\partial\Phi/\partial\phi|_{\phi=0}=0$ and $\dot{J}_i=0$. This is the requirement
introduced in Section~\ref{Sec::generatingfn} for the Fourier components of
the generating function to be real. Now let us consider the point $\btheta=(0,0,0)$. Here the particle is at pericentre, crossing the $z=0$ plane, and at $\phi=0$. At this point we have
\begin{equation}
\frac{\partial r}{\partial \theta_i} = \cos\vartheta = p_r = \frac{\partial p_\vartheta}{\partial \theta_i} = \frac{\partial p_\phi}{\partial \theta_i} = 0,
\end{equation}
so
\begin{equation}
\dot{J}_i = -\frac{\partial\Phi}{\partial\phi}\frac{\partial\phi}{\partial\theta_i}-\frac{\partial\Phi}{\partial\vartheta}\frac{\partial\vartheta}{\partial\theta_i}.
\end{equation}
As we saw before the first term is zero as $x=0$ is a symmetry plane of the potential. The second term is also zero as $z=0$ is also a symmetry plane. By a similar argument at
$\btheta=(0,\pi/2,0)$, $\partial\Phi/\partial\phi|_{\phi=\pi/2}=0$ as $y=0$ is
a symmetry plane of the potential. 

We calculate $\dot{\bs{J}}$ from equation~\eqref{toyact} as  
\begin{equation}
\dot{\bs{J}} = \sum_{\bs{n}\in\mat{N}}2\bs{n}\Big(\i \bs{n}\cdot\dot{\btheta}\Big)S_{\bs{n}}(\bs{J}')\sin\bs{n}\cdot\btheta
\end{equation}
At the point $\btheta=(0,0,\pi/2)$ we know $\dot{\bs{J}}=\bs{0}$ so we
require $\sin\pi n_3/2=0$ so $n_3$ must be even. Similarly we know
$\dot{\bs{J}}=\bs{0}$ at $\btheta=(0,\pi/2,0)$ so $n_2$ is restricted to
even values. However, $n_1$ can take any integer value.

\subsection{Box orbits}
Now let us consider the box orbits. We have a target Hamiltonian of the form
\begin{equation}
H = \frac{1}{2}\sum_i p_i^2+\Phi(x,y,z)
\end{equation}
where $p_i=(p_x,p_y,p_z)$ and the equations of motion for the toy actions are
\begin{equation}
\dot{J}_i = -\sum_j\frac{\partial\Phi}{\partial x_j}\frac{\partial x_j}{\partial \theta_i}-p_j\frac{\partial p_j}{\partial \theta_i}
\end{equation}
Consider the point $\btheta=(0,0,0)$. Here the orbit is turning in all three coordinates so $\bs{p}=\bs{0}$ and $\partial \bs{x}/\partial \theta_i=\bs{0}$ so $\dot{\bs{J}}=\bs{0}$ as required in Section~\ref{Sec::generatingfn}. Now let us consider the point $\btheta=(\pi/2,0,0)$. Here the orbit is turning in $y$ and $z$ and is passing through the $x=0$ plane at which point $\partial p_x/\partial \theta_i=0$ as $p_x$ is at a maximum. Therefore we have 
\begin{equation}
\dot{J}_i = -\frac{\partial\Phi}{\partial x}\frac{\partial x}{\partial \theta_i}.
\end{equation}
 For a triaxial potential aligned with our choice of Cartesian axes $x=0$ is
a symmetry plane so $\partial\Phi/\partial x |_{x=0} = 0$. Therefore
$\dot{\bs{J}}=\bs{0}$ here and by similar arguments to the loop orbit case we
are restricted to even $n_1$. We can employ the same arguments by considering
the stationary points $\btheta=(0,\pi/2,0)$ and $\btheta=(0,0,\pi/2)$
to show that $n_2$ and $n_3$ must be even.

\section{Angles and frequencies}\label{Appendix::Angles}
To find the angles and frequencies from an orbit timeseries we must minimize equation~\eqref{eq::angpenalty} with respect to the unknowns. The unknowns are $\btheta'(0)$, $\bs{\Omega}'$ and the set of $\partial S_{\bs{n}}/\partial \bs{J}'$, which we denote as $(\partial_1S_{\bs{n}},\partial_2S_{\bs{n}},\partial_3S_{\bs{n}})$. 
For each time we define the $N$-vector 
\begin{equation}
\mat{s}_{\bs{n}}(t_i) = -2\sin\bigl(\bs{n}\cdot\btheta(t_i)\bigr).
\end{equation}
We also define the $3(2+N)$-vectors
\begin{equation}
\bs{x}_{\btheta}\equiv
(\btheta'(0),\bs{\Omega}',\partial_1S_{\bs{n}},\partial_2S_{\bs{n}},\partial_3S_{\bs{n}}),
\end{equation}
\begin{equation}
\bs{b}_{\btheta}\equiv \sum_i(\btheta(t_i),t_i \btheta(t_i), \theta_1(t_i)\mat{s}(t_i),\theta_2(t_i)\mat{s}(t_i),\theta_3(t_i)\mat{s}(t_i))
\end{equation}
and the symmetric matrix
\begin{equation}
\mat{A}_{\btheta}\equiv \sum_i
\left(
\begin{array}{ccccc}
  \mat{I}_3 & t_i\mat{I}_3 & \mat{s}^{1\T} &\mat{s}^{2\T}&\mat{s}^{3\T}\\
  t_i\mat{I}_3 & t^2_i\mat{I}_3 & t_i\mat{s}^{1\T} &t_i\mat{s}^{2\T}&t_i\mat{s}^{3\T}\\
  \mat{s}^{1} & t_i\mat{s}^{1} & \mat{s}\cdot\mat{s}^\T & 0 & 0\\
  \mat{s}^{2} & t_i\mat{s}^{2} & 0 & \mat{s}\cdot\mat{s}^\T & 0\\
  \mat{s}^{3} & t_i\mat{s}^{3} & 0 & 0 & \mat{s}\cdot\mat{s}^\T\\
\end{array}
\right),
\end{equation}
where each $\mat{s}^{m}$ is an $N$-by-$3$ matrix with the $N$-vector
$\mat{s}$ in the $m$th column, and each $\mat{s}$ is evaluated at the $i$th time.
Setting the partial derivatives of $F$ with respect to the unknowns to zero yields the matrix equation,
\begin{equation}
\mat{A}_{\btheta}\cdot\bs{x}_{\btheta}=\bs{b}_{\btheta}.
\end{equation}

\bsp

\label{lastpage}
\end{document}